\documentclass[twocolumn]{revtex4}
\usepackage{graphicx}
\usepackage{bm}
\usepackage{color}
\usepackage{amsmath}
\usepackage{natbib}

\begin{document}

\preprint{APS/123-QED}

\title{Dispersion properties of transverse waves in electrically polarized BECs}

\author{Pavel A. Andreev}
\email{andreevpa@physics.msu.ru}
\author{L. S. Kuz'menkov}%
\email{lsk@phys.msu.ru}
\affiliation{M. V. Lomonosov Moscow State University, Moscow, Russian Federation.}

\date{\today}

\begin{abstract}
Further development of the method of quantum hydrodynamics
in application for Bose-Einstein condensates (BECs) is presented.
To consider evolution of polarization direction along with
particles movement we have developed corresponding set of quantum
hydrodynamic equations. It includes equations of the
polarization evolution and the polarization current evolution
along with the continuity equation and the Euler equation
(the momentum balance equation). Dispersion properties of the transverse waves including the
electromagnetic waves propagating through the BECs are considered.
To this end we consider full set of the Maxwell equations for description of electromagnetic field
dynamics. This approximation gives us possibility to
consider the electromagnetic waves along with the matter waves. We find a splitting of the
electromagnetic wave dispersion on two branches. As a result we have four solutions, two
for the electromagnetic waves and two for the matter waves, the last two
are the concentration-polarization waves appearing as a generalization of the Bogoliubov mode. We also obtain that if the matter wave propagate perpendicular to external electric field when dipolar contribution does not disappear (as it follows from our generalization of the Bogoliubov spectrum). In this case exist a small dipolar frequency shift due to transverse electric field of perturbation.
\end{abstract}

\pacs{03.75.Kk  03.75.Hh}
\keywords{Keywords: Bose-Einstein condensate; collective
excitations; polarization; transverse waves; quantum hydrodynamic
model}
\maketitle


\section{\label{sec:level1} Introduction}

At studying of electrically polarized Bose-Einstein condensates
(BECs) the generalization of the Gross-Pitaevskii (GP) equation
\cite{Goral PRA 00}-\cite{Fischer PRA 06R} and some its further
generalization \cite{Wang NJP 08}-\cite{Lima PRA 12} have been used, see
also review papers \cite{Lahaye RPP 09}-\cite{Baranov CR 12}. This nonlinear
Schrodinger equation defines the complex scalar wave function
$\Phi(\textbf{r},t)$, which describes evolution of the particle concentration
$n(\textbf{r},t)=\Phi^{*}(\textbf{r},t)\Phi(\textbf{r},t)$. When
direction of polarization of each particle $\textbf{d}$ is not
changing during the considering processes and all of them have same
direction (the aligned dipoles) we can write formula for the density of polarization
$\textbf{P}=\Phi^{*}(\textbf{r},t)\textbf{d}\Phi(\textbf{r},t)$.
Thus, the polarization density changes due to particles movement, and
consequently, at changing of concentration. The waves of polarization
can easily propagate even in the ferroelectric materials as well as the spin
waves in ferromagnetic materials. All the more we can expect existence of
such waves in the polarized BECs. To consider evolution of polarization
direction along with particles movement we have developed the set
of quantum hydrodynamic equations, which includes equations of the
polarization evolution and the polarization current evolution
along with the continuity equation and the Euler equation
(the momentum balance equation) \cite{Andreev arxiv 12
02}-\cite{Andreev PRB 11}. Earlier we developed similar approach for the electric dipolar ultracold fermions \cite{Andreev RPJ 13}.

There is a field in dipolar BECs studying, where evolution of dipole directions is under consideration. This is research of spinor BECs, which was suggested by T. Ohmi and K. Machida \cite{Machida JPSJ 98} and T. L. Ho \cite{Ho PRL 98} in 1998 and was recently reviewed in Refs. \cite{Kawaguchi Ph Rep 12}, \cite{Stamper-Kurn RMP 13}. We should mention that this field is dedicated to magnetic dipole BECs, but the dipole-dipole (spin-spin) interaction is neglected there.

It was marked in Ref. \cite{Andreev arxiv 12 02} that the full and correct form of Hamiltonian of the electric dipole interaction, which also corresponds to the Maxwell equations, is
\begin{equation}\label{di BEC tran Ham dd int} H_{dd}=-\partial^{\alpha}\partial^{\beta}\frac{1}{r}\cdot d_{1}^{\alpha}d_{2}^{\beta}.\end{equation}
Justification of this formula in accordance with classical electrodynamics \cite{Landau 2} is presented in the Appendix A.

Using well-known identity (see for instance Refs. \cite{O'Dell PRL 04}, \cite{Bao JCP 10})
\begin{equation}\label{di BEC tran togdestvo} -\partial^{\alpha}\partial^{\beta}\frac{1}{r}= \frac{\delta^{\alpha\beta}-3r^{\alpha}r^{\beta}/r^{2}}{r^{3}}+\frac{4\pi}{3}\delta^{\alpha\beta}\delta(\textbf{r}),\end{equation}
we can see that the Hamiltonian (\ref{di BEC tran Ham dd int}) differs from usually used one
\begin{equation}\label{di BEC tran pot of dd int usual} H_{dd}=\frac{\delta^{\alpha\beta}-3r^{\alpha}r^{\beta}/r^{2}}{r^{3}}d_{1}^{\alpha}d_{2}^{\beta}.\end{equation}
Moreover, the Hamiltonian (\ref{di BEC tran Ham dd int}) is the original form of the potential energy of electric dipole interaction.
Necessity to consider the Hamiltonian of electric dipole interaction in the form (\ref{di BEC tran Ham dd int}) caused by the fact that it must accord to the Maxwell equations. This connection exists since Maxwell equations describe the electric field and its connection with the sources. In our case source of the field is the density of electric polarization. Action of the electric field on the density of polarization comes in to equations of motion via the force field. As the result it describes interaction of the polarization (the electric dipole moments).

It has been shown in Refs. \cite{Santos PRL 03}-\cite{Giovanazzi
EPJD 04} that using of the non-linear Schrodinger equation,
which generalized the GP equation for the electrically polarized BECs, and using of the formula (\ref{di BEC
tran pot of dd int usual}) for description of the dipole-dipole
interaction lead to the appearance of the contribution of the
equilibrium polarization in the dispersion dependence of the
Bogoliubov mode. This contribution leads to the anisotropy of
the dispersion dependence. The anisotropy of the sound velocity in dipolar Bose gases has recently been confirmed experimentally (see Ref. \cite{Bismut PRL 12}).

In Refs. \cite{Andreev arxiv Pol}, \cite{Andreev RPJ 12} we considered the evolution of electric dipole directions
and got dispersion dependencies for two waves. That means what
account of polarization evolution leads to appearing of new wave
in the BECs.  This is a polarization wave, which is an analog of spin waves in systems of spinning particles.

At the first step of studying of the polarization direction evolution we considered propagation of waves parallel to direction of electric field, so their dispersion reveals no angle dependence \cite{Andreev arxiv Pol}, \cite{Andreev RPJ 12}. As an intermediate step in studying of the polarization evolution in BECs we considered fully polarized BECs.
Usual theory of fully polarized BECs is based on formula (\ref{di BEC tran pot of dd int usual}) (see for instance \cite{Lahaye RPP 09}, \cite{Baranov P R 08}, \cite{Baranov CR 12}). It has the following consequences.
Dipolar part of $\omega^{2}$ is proportional to $(\cos^{2}\theta-1/3)$, so it can change sign, where $\theta$ is the angle between the direction of wave propagation $\textbf{k}$ and the direction of external field. If the polarization part is negative at positive and small enough short-range part of the spectrum, one finds the roton instability \cite{Lahaye RPP 09}, \cite{Baranov P R 08}, \cite{Baranov CR 12}.
Our consideration of fully polarized electric dipolar BECs is based on formula (\ref{di BEC tran Ham dd int}). Consequently the dipolar part of $\omega^{2}$ is proportional $\cos^{2}\theta$. So it is positive for all angles $\theta$ and shows no instability of three dimensional BECs with the repulsive short range interaction \cite{Andreev 1401.4842}. Moreover full electric dipole interaction (\ref{di BEC tran Ham dd int}) leads to stabilization of the Bogoliubov spectrum with attractive short range interaction \cite{Andreev 1401.4842}, since positive dipolar part can exceed negative contribution of the short range interaction.
We obtained that consideration of the full Hamiltonian (\ref{di BEC tran Ham dd int}) is essential for fully polarized electric BECs, the full Hamiltonian of spin-spin interaction was considered for the fully polarized magnetic BECs as well. In both cases we have one wave solution only, as it was presented in Refs. \cite{Andreev 1401.4842}. However the dipolar part of dispersion dependence changes due to account of the delta-function part of the full potential. As a result we find that the dipolar part in the electric dipolar BECs becomes positive for all angles (all directions of wave propagation). We have different picture for the magnetic dipolar BECs, where we have another coefficient in front of the delta-function (see formula (83.9) in Ref. \cite{LL 4}). Consequently the dipolar part of spectrum is also different. It is negative for all angles. Hence, there is no roton instability for the electrically polarized BECs, but the roton instability exists in the magnetically polarized BECs at all angles for small enough constant of the short-range interaction.

Consideration of the full electric dipolar BEC theory including the dipoles direction evolution was presented in Ref. \cite{Andreev arxiv 12 02}, where we applied our model to calculation of dispersion of waves propagating at different angles $\theta$. Thus Ref. \cite{Andreev arxiv 12 02} contains generalization of results obtained in Ref. \cite{Andreev arxiv Pol}. Waves considered in Refs. \cite{Andreev arxiv 12 02}, \cite{Andreev arxiv Pol}, \cite{Andreev RPJ 12} were longitudinal, that means perturbations of the electric field is parallel to the direction of wave propagation $\textbf{k}\parallel \textbf{E}$. Many features of the quantum hydrodynamics of dipolar BECs and meaning of considered approximations were described in Ref. \cite{Andreev arxiv 12 02} as well.

In this paper we suggest more general model in comparison with our recent papers. We consider full set of the Maxwell equations instead of the pair quasi-electrostatic equations describing the electric field considered earlier. This approximation gives us possibility to consider electromagnetic waves along with the matter waves. Below we will show that in the electrically polarized BECs we have a splitting of the electromagnetic waves on two branches. So, we have four solution, two for the electromagnetic waves and two for the matter waves, the last two are waves of concentration-polarization. It is important and interestingly to admit that in this approximation we find anisotropy in the dispersion dependencies for all four waves.

This paper is organized as follows. In Sec. II we present the quantum hydrodynamics model for description of the electrically polarized BECs. In Sec. III we present dispersion equation and describe the method to get it. In Sec. IV we present analysis of the dispersion equation.  In Sec. V brief discussion of obtained results is presented.

\section{\label{sec:level1} Basic equations}

In our previous paper  \cite{Andreev arxiv 12 02}, \cite{Andreev 2013 non-int GP} we have presented the derivation of the QHD equation for the electrically polarized BECs. These equations were directly derived from the microscopic many-particle Schrodinger equation by means the QHD method. Method of quantum hydrodynamics was suggested in 1999 for derivation of hydrodynamic equations for quantum plasmas \cite{MaksimovTMP 1999}. In 2008 this method was applied to ultracold quantum gases \cite{Andreev PRA08}. Later it was used for system of particles baring electric dipole moment \cite{Andreev arxiv 12 02}, \cite{Andreev RPJ 12}, \cite{Andreev PRB 11}. Let us now present the set of the quantum hydrodynamics equations.

The first equation of the QHD equations system is the continuity
equation
\begin{equation}\label{di BEC tran cont eq}\partial_{t}n+\partial^{\alpha}(nv^{\alpha})=0\end{equation}
showing conservation of the particle number, and giving time evolution of the particle concentration.

The momentum balance equation (Euler equation) for the polarized BECs has the form
$$mn(\partial_{t}+\textbf{v}\nabla)v^{\alpha}+\partial_{\beta}p^{\alpha\beta}$$
$$-\frac{\hbar^{2}}{4m}\partial^{\alpha}\triangle
n+\frac{\hbar^{2}}{4m}\partial^{\beta}\Biggl(\frac{\partial^{\alpha}n\cdot\partial^{\beta}n}{n}\Biggr)
$$
\begin{equation}\label{di BEC tran bal imp eq short}=\Upsilon n\partial^{\alpha}n+\frac{1}{2}\Upsilon_{2}\partial^{\alpha}\triangle n^{2}+P^{\beta}\partial^{\alpha}E^{\beta},
\end{equation}
where
\begin{equation}\label{di BEC tran Upsilon} \Upsilon=\frac{4\pi}{3}\int
dr(r)^{3}\frac{\partial U(r)}{\partial r},
\end{equation}
and
\begin{equation}\label{di BEC tran Upsilon2} \Upsilon_{2}\equiv\frac{\pi}{30}\int dr
(r)^{5}\frac{\partial U(r)}{\partial r},\end{equation}
are the numerical coefficients. In equation (\ref{di BEC tran bal imp eq short})  we define a
parameter $\Upsilon_{2}$ as (\ref{di BEC tran Upsilon2}). This
definition differs from the one in the work \cite{Andreev PRA08}. Here we put multiplier $1/8$ within the definition of $\Upsilon_{2}$ to make equations looks better.
Terms proportional to $\hbar^{2}$ appear as a result of usage of the
quantum kinematics. They are usually called the quantum Bohm potential. The first two terms at the right-hand side of the
equation (\ref{di BEC tran bal imp eq short}) are first terms of
expansion of the quantum stress tensor. They occur due to account of the short-range interaction potential $U_{ij}$. The interaction
potential $U_{ij}$ determinates the macroscopic interaction
constants $\Upsilon$ and $\Upsilon_{2}$. Term proportional $\Upsilon_{2}$ contains higher spatial derivation in compare with the Gross-Pitaevskii term $\nabla n^{2}/2$. This is why it is called a non-local interaction. Contributions of the third order by the radius of the short range interaction (terms proportional $\Upsilon_{2}$) in non-linear properties of quantum gases were considered in Refs. \cite{Andreev MPL B 12}, \cite{Zezyulin EPJ D 13} for non-polarized quantum gases. Influence of $\Upsilon_{2}$ on dispersion of waves in dipolar fermions was described in Ref. \cite{Andreev RPJ 13}. A non-local interaction for BECs being used as an analogue gravity system in Ref. \cite{Sarkar ariv 13}. The last term of
the equation (\ref{di BEC tran bal imp eq short}) describes force field
that affects the dipole moment in a unit of volume as the effect of
the external electrical field and the field produced by other
dipoles. It is written using the
self-consistent field approximation \cite{Andreev arxiv 12 02},
\cite{Andreev PRB 11}, \cite{MaksimovTMP 1999}. $p^{\alpha\beta}(\textbf{r},t)$ is the
tensor of the kinetic pressure, which depends on particle
thermal velocities and makes no contribution into the BEC dynamics
at near zero temperatures. In equation (\ref{di BEC tran bal imp eq short}) and below $\textbf{E}$ is the sum of external and internal electric fields, where internal electric field is created by dipoles. $\textbf{P}$ is the density of electric dipole moment (the polarization of medium).

The first order by the interaction radius constant for
dilute gases may be presented in the form
\begin{equation}\label{di BEC tran Upsilon via scattering length} \Upsilon=-\frac{4\pi\hbar^{2}a}{m},\end{equation}
where $a$ is the scattering length, so we see $\Upsilon=-g$, where $g$ is the Gross-Pitaevskii interaction constant \cite{L.P.Pitaevskii RMP
99,Andreev PRA08}.

We have also the field equations
\begin{equation}\label{di BEC tran field eq div bas}\nabla\textbf{E}(\textbf{r},t)=-4\pi \nabla\textbf{P}(\textbf{r},t),\end{equation}
and
\begin{equation}\label{di BEC tran field eq rot bas}\nabla\times\textbf{E}(\textbf{r},t)=0.\end{equation}
These equations show relation between the polarization $\textbf{P}$ and the electric field $\textbf{E}$ caused by the polarization $\textbf{P}$.
Equation (\ref{di BEC tran field eq div bas}) allows to consider propagation of waves parallel to the equilibrium polarization only. For consideration of waves propagating in arbitrary direction we have to consider the couple of equations (\ref{di BEC tran field eq div bas}) and (\ref{di BEC tran field eq rot bas}).

Having a short-range interaction in microscopic many-particle Schrodinger equation we obtain corresponding force field $\textbf{F}_{SR}$ in the Euler equation as result of application of the QHD method for derivation of equations for collective evolution description \cite{Andreev PRA08}. Using the fact that interaction is short-range we find the force field $\textbf{F}_{SR}$ as a series. The force field also appears as the divergence of a second rank tensor, which is the quantum stress tensor $\sigma^{\alpha\beta}$. Considering the first two non-zero terms in the series, existing for spherically symmetric potential of the short-range interaction for particles in the BEC state, we get the Gross-Pitaevskii term and a non-local term containing the third spatial derivative of the square of the particle concentration $\textbf{F}_{SR}^{\alpha}=-\partial_{\beta}\sigma^{\alpha\beta}=-\frac{1}{2}g\nabla^{\alpha} n^{2}+\frac{1}{2}\Upsilon_{2}\nabla^{\alpha}\triangle n^{2}$, with $\sigma^{\alpha\beta}=\frac{1}{2}g\delta^{\alpha\beta}n^{2}-\frac{1}{2}\Upsilon_{2}\partial^{\alpha}\partial^{\beta}n^{2}$. As a consequence we see that there is no corresponding non-linear Schrodinger equation, since there is no Cauchy integral of the Euler equation. Nevertheless a formal representation of the hydrodynamic equations in form of a non-linear Schrodinger equation was made in Ref. \cite{Andreev PRA08}.

One may try to include non-local terms in the GP equation or the corresponding Lagrangian including spatial derivatives of the particle concentration. For instance one may consider $gn-\Upsilon_{2}\triangle n$ instead of $gn$. However, a microscopic derivation of QHD equations shows that the Euler equation has more complicate form. We found $\textbf{F}=-\frac{1}{2}g\nabla n^{2}+\frac{1}{2}\Upsilon_{2}\nabla\triangle n^{2}$ instead of $\textbf{F}=-n\nabla(g-\Upsilon_{2}\triangle) n$ corresponding to a non-linear Schrodinger equation.

If particles do not contain the dipole moment, the
continuity equation and the momentum balance equation form a
closed set of equations. If the dipole moment is taken into
account in the momentum balance equation when a new physical quantity
emerges. This is the polarization vector field $\textbf{P}(\textbf{r},t)$.
It makes the set of equations to become incomplete.

\begin{figure}
\includegraphics[width=8cm,angle=0]{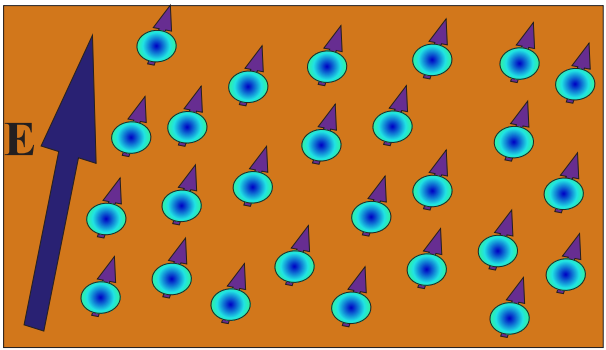}
\caption{\label{di BEC tran 00_arrows_parallel} (Color online) This figure shows motion of fully polarized dipoles, when direction of all dipoles does not change. Hence the polarization of the system change in accordance with change of the particle concentration.}
\end{figure}

If we have a system of fully polarized particles pictured on Fig. (\ref{di BEC tran 00_arrows_parallel}) when polarization changes due to motion of particles in space. Consequently we have $\textbf{P}(\textbf{r},t)=d \textbf{l} n(\textbf{r},t)$, where $d$ is the electric dipole moment of particles, $\textbf{l}$ is the direction of all dipoles. A non-linear Schrodinger equation can be derived for system of fully polarized dipoles from the first principal derived QHD equations \cite{Andreev PRA08}. In an integral form it can be written as
$$\imath\hbar\partial_{t}\Phi(\textbf{r},t)=\biggl(-\frac{\hbar^{2}\nabla^{2}}{2m}+\mu(\textbf{r},t)+V_{ext}(\textbf{r},t)+g\mid\Phi(\textbf{r},t)\mid^{2}$$
\begin{equation}\label{di BEC tran GP eq int form}+d^{2}\int d\textbf{r}' \biggl[
\frac{1-3\cos^{2}\widetilde{\theta}}{|\textbf{r}-\textbf{r}'|^{3}}\mid+\frac{4\pi}{3}\delta(\textbf{r}-\textbf{r}')\biggr]\Phi(\textbf{r}',t)\mid^{2}\biggr)\Phi(\textbf{r},t),
\end{equation}
with $\Phi(\textbf{r},t)$ is the macroscopic wave function, $\mu$ is the chemical potential, $V_{ext}$ is the potential of an external field, $g$ is the constant of short-range interaction, $d$ is the dipole electric moment of a single particle, $\widetilde{\theta}$ is the angle between $\textbf{r}-\textbf{r}'$ and direction of external electric field, $m$ is the mass of a particle and $\hbar$ is the Planck constant divided by $2\pi$.
This is a generalized GP equation, where we included the Dirac delta function part of the full potential of electric dipole interaction, as we did earlier in Appendix B of Ref. \cite{Andreev arxiv 12 02}. Let us admit that the macroscopic wave function $\Phi(\textbf{r},t)$ appears to be defined in hydrodynamic variables as follows $\Phi(\textbf{r},t)=\sqrt{n(\textbf{r},t)}\exp(\imath m\phi(\textbf{r},t)/\hbar)$ with $\phi(\textbf{r},t)$ is the potential of the velocity field $\textbf{v}(\textbf{r},t)=\nabla\phi(\textbf{r},t)$.

\begin{figure}
\includegraphics[width=8cm,angle=0]{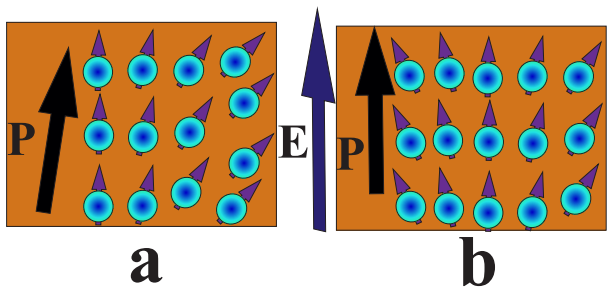}
\caption{\label{di BEC tran 00_arrows and polarization} (Color online) This figure shows formation of full polarization in a vicinity of a point at dipole direction evolution. Fig. \ref{di BEC tran 00_arrows and polarization}a illustrates possibility of turn of the polarization vector $\textbf{P}$ due to asymmetrical distribution of dipole directions, which can appear at propagation of the polarization wave. Fig. \ref{di BEC tran 00_arrows and polarization} shows decreasing of module of the polarization $\textbf{P}$ without change of direction. It happens due to symmetric distribution of dipole direction in the vicinity.}
\end{figure}

The Delta function term in equation (\ref{di BEC tran GP eq int form}) can be combined with the short range interaction term presented by the term with the cubic nonlinearity. So we might obtain shifted constant of the short range interaction $\tilde{g}=g+4\pi/3$. But this is unwise move. We can go other way allowing to simplify equation (\ref{di BEC tran GP eq int form}) without extra approximations. We can explicitly introduce the electric field created by dipoles (see Appendix B formula (\ref{di BEC tran el field of dip explict})) and rewrite the non-linear Schrodinger equation in a non-integral form \cite{Andreev 2013 non-int GP}
\begin{equation}\label{di BEC tran GP non Int} \imath\hbar\partial_{t}\Phi(\textbf{r},t) =\Biggl(-\frac{\hbar^{2}}{2m}\triangle+g\mid\Phi(\textbf{r},t)\mid^{2}-\textbf{d}\textbf{E}\Biggr)\Phi(\textbf{r},t).\end{equation}
We do not included the chemical potential in equation (\ref{di BEC tran GP non Int}). Equation (\ref{di BEC tran GP non Int}) contains an extra function $\textbf{E}(\textbf{r},t)$, which obey the quasi electrostatic Maxwell equations (\ref{di BEC tran field eq div bas}) and (\ref{di BEC tran field eq rot bas}). So we have a set of coupled equation for medium motion (\ref{di BEC tran GP non Int}) and field (\ref{di BEC tran field eq div bas}) and (\ref{di BEC tran field eq rot bas}). The non-integral form (\ref{di BEC tran GP non Int}) of the generalized GP equation (\ref{di BEC tran GP eq int form}) was introduced in Ref. \cite{Andreev 2013 non-int GP}. See also discussion of this equation in Ref. \cite{Andreev 1401.4842}. Let us admit that in this paper we discuss theory of dipolar BECs in model of point-like particles. Finite size particles were considered in Ref. \cite{Andreev 1401.4842}, where a radius of molecules explicitly comes in the model.

The short range interaction constant $g$ in equation (\ref{di BEC tran GP non Int}) does not depend on electric dipole moment $d$. Hence the corresponding scattering length does not depend on $d$ either.

The dipole dependent scattering length for the short range part of interaction in dipolar BECs is under discussion in Ref. \cite{Bortolotti PRL 06}. Authors consider a potential, which is close to the real one, with simplified treatment of the short range interaction, which considered as infinite repulsion at a cutoff radius (see formula (2) of Ref. \cite{Bortolotti PRL 06}). Considering GP equation they pay attention to an appropriate pseudopotential, its part proportional to the delta function is considered to be dependent on the dipole moment. So they have a dipole dependent scattering length. Due to numerical analysis it is hard to get an appropriate picture of dependence of the scattering length on the dipole moment and the cutoff radius. Nevertheless consideration of a close to realistic potential is related to a theory of finite size dipolar particles presented in \cite{Andreev 1401.4842}, where dependence on particle finite size and dipole moment is presented explicitly and analytically. No pseudopotential was used in Ref. \cite{Andreev 1401.4842} applying the quantum hydrodynamic method. Application of the dipole dependent scattering length $\tilde{a}(d)$ along with the no cutoff of the dipole-dipole potential for finite particles means that $\tilde{a}(d)$ contains a contribution aimed to cancel an endowment of dipole-dipole interaction inside of the sphere with the cutoff radius. $\tilde{a}(d)$ contains dependence on cutoff radius as well.

Refs. \cite{Bortolotti PRL 06} and \cite{Kanjilal PRA 07} presented $\tilde{a}(d)$ appearing at consideration of dipolar BECs in terms of the pseudopotential. This dependence is mentioned in reviews on dipolar BECs (see for instance formula (3) of Ref. \cite{Baranov P R 08} and text after it, and formula (4.6) of Ref. \cite{Lahaye RPP 09} and its discussion below). Unfortunately dipole dependence of the interaction constant $\tilde{g}(d)=4\pi\hbar^{2}\tilde{a}(d)/m$ is not included in discussions of spectrum of bulk collective Bogoliubov excitations. These discussions (see Ref. \cite{Baranov P R 08} text after formula (6), \cite{Baranov CR 12} text after formula (11), \cite{Lahaye RPP 09} subsection (5.1)) of dipole dependence of the Bogoliubov spectrum happens as there is no dipole dependent scattering length $\tilde{a}(d)$, but usual scattering length of dipoleless particles.

Getting back to our analysis we separated the short range interaction and the dipole-dipole interaction with the beginning of our paper. So we have the scattering length $a$, which does not depend on $d$. All dependencies on the electric dipole moment are included in the dipole-dipole interaction.

Generalized integral GP equation, but with the shorted potential of dipole-dipole interaction (\ref{di BEC tran pot of dd int usual}), still finds a lot of applications. Frequencies of collective excitations in trapped dipolar BECs were calculated
in the Thomas-Fermi regime in Ref. \cite{van Bijnen PRA 10}. One-dimensional solitons
on a weak two-dimensional square and triangular optical lattice potentials placed perpendicular to the
polarization direction were considered in Ref. \cite{Adhikari PL A 12}. Order and chaos in the dipolar BECs are investigated in Ref. \cite{Koberle NJP 09} considering behavior of bifurcations. A dipolar boson-fermion
mixture and its stability was studied in Ref. \cite{Adhikari PRA 13}. When stability of dipolar Bose and Fermi gases was investigated in Ref. \cite{Adhikari JP B 13}. Solitons in dipolar boson-boson mixtures were described in Ref. \cite{Adhikari JP B 14}. Influence of dipole-dipole interaction on the formation of vortices in a rotating dipolar BECs atoms in quasi
two-dimensional geometry was considered in Ref. \cite{Kumar JP B 12}. Solitons in dipolar BECs trapped in the two dimensional plane were studied in Ref. \cite{Kumar JP B 13}.  Modification of the properties of super-fluid vortices by dipole-dipole interactions within the context of a two-dimensional atomic Bose gas of fully polarized dipoles was considered in Ref. \cite{Mulkerin PRL}. A discussion of linear properties of dipolar BEC based on integral GP equation with shorted dipole-dipole potential (\ref{di BEC tran pot of dd int usual}) is also presented in Ref. \cite{Mulkerin PRL}. Judging upon formula (2) in Ref. \cite{Qiuzi Li PRB 11} we assume that potential (\ref{di BEC tran pot of dd int usual}) was used for consideration of collective modes in fermionic dipolar liquid.

We can also find explicit examples of using of identity (\ref{di BEC tran togdestvo}) when authors work with the shorted potential of dipole-dipole interaction (\ref{di BEC tran pot of dd int usual}) see text before formula (418) in review \cite{Kawaguchi Ph Rep 12}, see formula (36) in Ref. \cite{Lazur PRA 10}, formula (8) in Ref. \cite{Eberlein PRA 05}, formula (8) and text before this formula in Ref. \cite{O'Dell PRL 04}, \textit{and} formula (4) in Ref. \cite{Cai PRA 10}.

\begin{figure}
\includegraphics[width=8cm,angle=0]{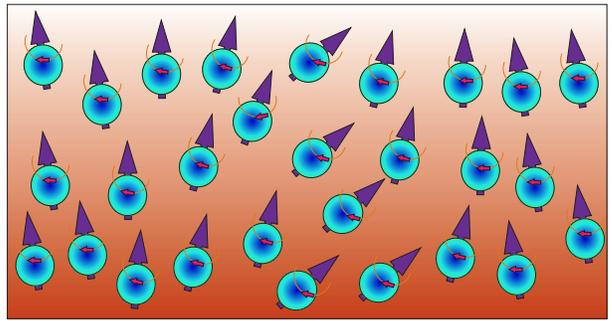}
\caption{\label{di BEC tran 01_arrows} (Color online) This figure shows motion of dipoles including evolution of dipole direction. Here we have two mechanisms of polarization change. The first of them is the same as in systems of fully polarized particles. It is motion of each particle in space, so the particle concentration changes and causes change of the polarization. The second mechanism is related to variation of direction of dipoles from direction of external field due to interparticle interaction.}
\end{figure}

Other approaches to dipolar BECs were developed in last years \cite{Lima PRA 12}, \cite{Lima PRA 10}, \cite{Babadi PRA 12}. Some of them we briefly describe below.
Various beyond mean-field effects on Bose gases at zero temperature were considered within the realm of the Bogoliubov-de Gennes theory \cite{Lima PRA 12}. For the homogeneous system, the condensate depletion, the ground-state energy, the equation of state, and the speed of sound are discussed in detail. Authors analyze the influence of quantum fluctuations on properties of quantum gases. During calculations the condensate density being replaced by the total number density, which includes distribution of particles on states with non minimal energy. The fluctuation Hamiltonian density is included. The corrected sound velocity due to beyond mean-field equation of state, which is a generalization of the Lee-Huang-Yang quantum corrected equation of state, is presented by formula (34). Corresponding hydrodynamic equations are presented by formulas (32) and (33) of Ref. \cite{Lima PRA 12}.
Applying the many-particle Schrodinger equation for ultracold dipolar Fermi gases in anisotropic traps, which are polarized in the z direction, the couple of quantum hydrodynamic equations, i.e. the continuity and Euler equations, was obtained via the equation of motion for the one-body density matrix \cite{Lima PRA 10}. Collective motion of polarized dipolar Fermi gases in the hydrodynamic regime was considered in Ref. \cite{Lima PRA 10R} applying a variational time-dependent Hartree-Fock approach.

Collective excitations of quasi-two-dimensional trapped dipolar fermions in framework of the collisional Boltzmann-Vlasov equation obtained by reducing the evolution equations of the nonequilibrium Green functions to quantum kinetic equations were considered in Ref. \cite{Babadi PRA 12}. This analysis is performed at finite temperatures. Transition from collisionless to hydrodynamic regime was considered.

In this paper we are interested in dipole direction evolution instead of consideration of fully polarized BECs. So let us make a step towards the main topic. The variation of dipole direction can be different in different areas of space. Consequently we have different magnitude of polarization in different points of space. This variations can cause a turning of the polarization vector $\textbf{P}$ relative to the direction of the external electric field $\textbf{E}_{ext}$ (see Fig. (\ref{di BEC tran 00_arrows and polarization} a)) or it can cause change of module of $\textbf{P}$ without change of direction (see Fig. (\ref{di BEC tran 00_arrows and polarization} b)).

If particles possessing electric dipole moment are in different cells of an optical lattice in the Mott insulator regime when they ability to get close to each other is limited. However there is no such limitation in the superfluid phase or when particles are in the absence of a lattice. So particles can get close to each other. In this case magnitude of the dipole-dipole interaction increases and direction of dipoles can change due to the dipole-dipole  interaction. It gives possibility for polarization wave propagation, which are waves related to evolution of dipole direction Fig. (\ref{di BEC tran 01_arrows}).

All of it engage us to include dipole direction evolution in our model. Consequently we derive next equation of the chain of QHD equations, which is the equation for time evolution of the polarization (density of the electric dipoles) $\textbf{P}$. Explicit definitions of the QHD variables allowing to derive the set of QHD equations are presented  in Appendix B.

Next equation we need for studying of the collective excitation dispersion is the
equation of polarization evolution
\begin{equation}\label{di BEC tran eq polarization} \partial_{t}P^{\alpha}(\textbf{r},t)+\partial^{\beta}R^{\alpha\beta}(\textbf{r},t)=0,\end{equation}
where $R^{\alpha\beta}(\textbf{r},t)$ is the current of polarization.

The equation (\ref{di BEC tran eq polarization}) does not contain
information about the effect of the interaction on the
polarization evolution. So we should consider equation for
$R^{\alpha\beta}(\textbf{r},t)$ evolution, which can be derived by means of the quantum hydrodynamic method \cite{Andreev PRA08}, \cite{Andreev PRB 11}.  Using the self-consistent field
approximation of the dipole-dipole interaction we obtain the
equation for the polarization current
$R^{\alpha\beta}(\textbf{r},t)$ evolution
$$\partial_{t}R^{\alpha\beta}+\partial^{\gamma}\biggl(R^{\alpha\beta}v^{\gamma}+R^{\alpha\gamma}v^{\beta}-P^{\alpha}v^{\beta}v^{\gamma}\biggr)$$
$$+\frac{1}{m}\partial^{\gamma}r^{\alpha\beta\gamma}-\frac{\hbar^{2}}{4m^{2}}\partial_{\beta}\triangle P^{\alpha}$$
$$+\frac{\hbar^{2}}{8m^{2}}\partial^{\gamma}\biggl(\frac{\partial_{\beta}P^{\alpha}\partial_{\gamma}n}{n}+\frac{\partial_{\gamma}P^{\alpha}\partial_{\beta}n}{n}\biggr)
$$
\begin{equation}\label{di BEC tran eq for pol current gen selfconsist
appr} =\frac{1}{2m}\Upsilon\partial^{\beta}\biggl(nP^{\alpha}\biggr)+\frac{\sigma}{m}\frac{P^{\alpha}P^{\gamma}}{n}\partial^{\beta}E^{\gamma}, \end{equation}
where $r^{\alpha\beta\gamma}(\textbf{r},t)$ presents the
contribution of the thermal movement of the polarized particles into the
dynamics of $R^{\alpha\beta}(\textbf{r},t)$. As we deal with the BECs
below, the contribution of $r^{\alpha\beta\gamma}(\textbf{r},t)$
may be neglected. The last term of the formula (\ref{di BEC tran
eq for pol current gen selfconsist appr}) includes both an external
electrical field and the self-consistent field that particle dipoles
create. This term contain numerical constant $\sigma$. The second group of terms in the left-hand side of equation (\ref{di BEC tran eq for pol current gen selfconsist appr}) has kinematic nature. It is an analog of $mn(\textbf{v}\nabla)\textbf{v}$ in the Euler equation (\ref{di BEC tran bal imp eq short}). The fourth and fifth group of terms are the contribution of the quantum Bohm potential in the polarization evolution. Their appearance is related to the de-Broglie wave nature of quantum particles. The first term in the right-hand side of equation (\ref{di BEC tran eq for pol current gen selfconsist appr}) describes the short range interaction. We restrict ourself with the first order by the interaction radius  approximation for the short range interaction in the equation for $R^{\alpha\beta}$ evolution. That corresponds the Gross-Pitaevskii approach to the short range interaction $\Upsilon=-g$.

The quasi-electrostatic Maxwell equations (\ref{di BEC tran field eq div bas}) and (\ref{di BEC tran field eq rot bas}) emerge in the considered non-relativistic limit and allow to study longitudinal waves only. When we use term longitudinal we assume that direction of the electric field perturbation is parallel to the direction of the wave propagation. This case has been considered in Refs. \cite{Andreev arxiv 12
02}-\cite{Andreev RPJ 12}. It is well-known that equations  (\ref{di BEC tran field eq div bas}) and (\ref{di BEC tran field eq rot bas}) are the part of the set of the Maxwell equations. Therefore, if we want to consider the transverse waves we should use the full set of the Maxwell equations, which has well-known form
\begin{equation}\label{di BEC tran M1} \nabla \textbf{B}(\textbf{r},t)=0 , \end{equation}
\begin{equation}\label{di BEC tran M0} \nabla\textbf{E}(\textbf{r},t)=-4\pi \nabla\textbf{P}(\textbf{r},t),\end{equation}
\begin{equation}\label{di BEC tran M2} \nabla\times\textbf{E}(\textbf{r},t)=-\frac{1}{c}\partial_{t}\textbf{B}(\textbf{r},t),\end{equation}
and
\begin{equation}\label{di BEC tran M3} \nabla\times\textbf{B}(\textbf{r},t)=\frac{1}{c}\partial_{t}\textbf{E}(\textbf{r},t)+\frac{4\pi}{c}\partial_{t}\textbf{P}(\textbf{r},t),\end{equation}
where $c$ is the speed of light, and $\textbf{B}$ is the vector of magnetic field. From equation (\ref{di BEC tran M2}) we see $curl \textbf{E}\neq0$ $\Rightarrow$ $\textbf{k}\times \textbf{E}\neq0$. Hence we have contribution of $\textbf{E}_{\perp}\perp \textbf{k}$ along with $\textbf{E}_{\parallel}\parallel \textbf{k}$. When $\textbf{E}_{\parallel}$ can be considered by means quasi electrostatic equations (\ref{di BEC tran field eq div bas}) and (\ref{di BEC tran field eq rot bas}) only. The fundamental fact that the time evolution of polarization leads to generation of the magnetic field is also included in described model by means of equation (\ref{di BEC tran M3}).

\section{\label{sec:level1} Collective excitations in the electric polarized BEC: Dispersion equation}

We can analyze the linear dynamics of collective excitations in the
polarized BEC using the QHD equations (\ref{di BEC tran cont eq}),
(\ref{di BEC tran bal imp eq short}),  (\ref{di
BEC tran eq polarization}), (\ref{di BEC tran eq for pol current gen
selfconsist appr}) and the Maxwell equations (\ref{di BEC tran M1})-(\ref{di BEC tran M3}). Let us assume the system is placed in an
external electrical field $\textbf{E}_{0}=E_{0}\textbf{e}_{z}$.
The values of concentration $n_{0}$ and polarization
$\textbf{P}_{0}=\kappa\textbf{E}_{0}$ for the system in an
equilibrium state are constant and uniform and equilibrium velocity field
$v^{\alpha}(\textbf{r},t)$, magnetic field $B^{\alpha}(\textbf{r},t)$ and tensor
$R^{\alpha\beta}(\textbf{r},t)$ equal zero.

We consider the small perturbation of equilibrium state
$$\begin{array}{ccc}n=n_{0}+\delta n,& v^{\alpha}=0+\delta v^{\alpha},& \end{array}$$
$$\begin{array}{ccc}E^{\alpha}=E_{0}^{\alpha}+\delta E^{\alpha},& B^{\alpha}=0+\delta B^{\alpha},& \end{array}$$
\begin{equation}\label{di BEC tran equlib state BEC}\begin{array}{ccc}& & P^{\alpha}=P_{0}^{\alpha}+\delta P^{\alpha}, R^{\alpha\beta}=0+\delta R^{\alpha\beta}.\end{array}\end{equation}
Substituting these relations into system of equations (\ref{di BEC tran cont eq}),
(\ref{di BEC tran bal imp eq short}),  (\ref{di
BEC tran eq polarization}), (\ref{di BEC tran eq for pol current gen
selfconsist appr}) and (\ref{di BEC tran M1})-(\ref{di BEC tran M3}) \textit{and} neglecting
nonlinear terms, we obtain a system of the linear homogeneous
equations in partial derivatives with constant coefficients.
Passing to the following representation for small perturbations
$\delta f$
$$\delta f =f(\omega, \textbf{k}) exp(-\imath\omega t+\imath \textbf{k}\textbf{r}) $$
yields the homogeneous system of algebraic equations. The electric
field strength is assumed to have a nonzero value. Expressing all
the quantities entering the system of equations in terms of the
electric field, we come to the equation
\begin{equation}\label{di BEC tran alg eq}\Lambda^{\alpha\beta}(\omega, \textbf{k})E^{\beta}(\omega, \textbf{k})=0,\end{equation}
where
$$\Lambda^{\alpha\beta}=\biggl(\frac{\omega^{2}}{c^{2}}-k^{2}\biggr)\delta^{\alpha\beta}+k^{\alpha}k^{\beta}$$
$$-4\pi\frac{\omega^{2}}{c^{2}}k^{2}\frac{P_{0}^{\alpha}P_{0}^{\beta}}{\omega^{2}-\frac{\hbar^{2}k^{4}}{4m^{2}}+\frac{\Upsilon k^{2}n_{0}}{2m}}\times$$
\begin{equation}\label{di BEC tran Lambda tensor}\times\biggl(\frac{\sigma}{m n_{0}}+\frac{\frac{k^{2} \Upsilon}{2m^{2}}}{\omega^{2}-\frac{\hbar^{2}k^{4}}{4m^{2}}+\frac{\Upsilon n_{0}k^{2}}{m}-\frac{\Upsilon_{2}k^{4}n_{0}}{m}}\biggr).\end{equation}
with $k^{2}=k^{2}_{x}+k^{2}_{y}+k^{2}_{z}$. $\Lambda^{\alpha\beta}$ is the dielectric permittivity tensor of electric dipolar BECs.

We suppose that the external electric field parallel to the z-axis.
Thereby, the equilibrium polarization $\textbf{P}_{0}$ also parallel
to the z-axis $\textbf{P}_{0}\parallel \textbf{e}_{z}$. In this case
tensor (\ref{di BEC tran Lambda tensor}) becomes more simple
$$\hat{\Lambda}(\omega,
\textbf{k})$$
\begin{equation}\label{di BEC tran Lambda matrix}
=\left(\begin{array}{ccc}\frac{\omega^{2}}{c^{2}}-k^{2}_{y}-k^{2}_{z}&
k_{x}k_{y}&
k_{x}k_{z}\\
k_{x}k_{y}& \frac{\omega^{2}}{c^{2}}-k^{2}_{x}-k^{2}_{z}&
k_{y}k_{z}\\
k_{x}k_{z} &
k_{y}k_{z}&
\frac{\omega^{2}}{c^{2}}\biggl(1+\beta(\omega)\biggr)-k^{2}_{x}-k^{2}_{y}\\
\end{array}\right),\end{equation}
where
$$\beta(\omega)\equiv -4\pi\frac{P_{0}^{2}}{n_{0}}\frac{k^{2}}{\omega^{2}-\frac{\hbar^{2}k^{4}}{4m^{2}}+\frac{\Upsilon k^{2}n_{0}}{2m}}\times$$
\begin{equation}\label{di BEC tran}\times\biggl(\frac{\sigma}{m }+\frac{k^{2} \Upsilon n_{0}}{2(m^{2}\omega^{2}-\hbar^{2}k^{4}/4+m\Upsilon n_{0}k^{2}-m\Upsilon_{2}k^{4}n_{0})}\biggr).\end{equation}

Dispersion equation emerges as condition of existence of nonzero perturbations of electric field. That means the determinant of matrix (\ref{di BEC tran Lambda matrix}) equals to zero: $det \hat{\Lambda}=0$. Using formula
$(\ref{di BEC tran Lambda matrix})$ we find an explicit form of the dispersion equation
$$(\omega^{2}-k^{2}c^{2})^{2}$$
\begin{equation}\label{di BEC tran disp eq}+\beta(\omega)\biggl(\omega^{4}-(1+\cos^{2}\theta)\omega^{2}k^{2}c^{2}+\cos^{2}\theta k^{4}c^{4}\biggr)=0.\end{equation}

\section{\label{sec:level1} Dispersion dependencies}

\subsection{Matter waves}

Equation (\ref{di BEC
tran disp eq}) simplifies in the low frequency limit ($\omega\ll kc$)
\begin{equation}\label{di BEC tran disp eq low fr gen} k^{2}c^{2}+\beta(\omega)\biggl(k^{2}c^{2}\cos^{2}\theta-\omega^{2}\biggr)=0. \end{equation}

If $\cos\theta$ is not close to zero, when the first term in the large brackets is mach more than the second one. Consequently equation (\ref{di BEC tran disp eq low fr gen}) simplifies to
\begin{equation}\label{di BEC tran disp eq low fr}1+ \cos^{2}\theta \beta(\omega)=0,\end{equation}
at $\cos^{2}\theta =1$ we have
\begin{equation}\label{di BEC tran disp eq low fr simple}1+ \beta(\omega)=0.\end{equation}
Equation (\ref{di BEC tran disp eq low fr simple}) was obtained in
Refs. \cite{Andreev arxiv 12 02}, \cite{Andreev arxiv Pol}, \cite{Andreev RPJ 12}. Equation (\ref{di BEC tran disp eq low fr}) was also obtained in Ref. \cite{Andreev arxiv 12 02}, its properties were considered under condition of maximal polarization contribution, that corresponds $\cos\theta=1$. So properties of equation (\ref{di BEC tran disp eq low fr simple}) were studied in full details in Ref. \cite{Andreev arxiv 12 02}.

Equation (\ref{di BEC tran disp eq low fr}) has following solution
$$\omega^{2}=\frac{1}{2m}\Biggl(-\frac{3}{2}\Upsilon n_{0}k^{2}+\frac{\hbar^{2}k^{4}}{2m}+\Upsilon_{2}n_{0}k^{4}$$
$$+ 4\pi\sigma\cos^{2}\theta \frac{P_{0}^{2}k^{2}}{n_{0}}\pm\Biggl[\biggl(\frac{1}{2}\Upsilon n_{0}k^{2}-\Upsilon_{2}n_{0}k^{4}$$
\begin{equation}\label{di BEC tran disp dep low FR anizotr evident form}+ 4\pi\sigma\cos^{2}\theta \frac{P_{0}^{2}k^{2}}{n_{0}}\biggr)^{2}-8\pi\Upsilon k^{4}\cos^{2}\theta P_{0}^{2}\Biggr]^{1/2}\Biggr).\end{equation}
In the case of wave propagation parallel to the equilibrium polarization $\cos\theta=1$ solution we get solution obtained in Refs. \cite{Andreev arxiv 12 02}, \cite{Andreev arxiv Pol}, \cite{Andreev RPJ 12}, and at great length considered in Ref. \cite{Andreev arxiv 12 02}. The contribution of equilibrium polarization becomes smaller at increasing of angle $\theta$ from $0$ to $\pi/2$ and vanishes at $\theta=\pi/2$. However equation (\ref{di BEC tran disp eq low fr}) does not describe area around $\theta=\pi/2$. This area is described by equation (\ref{di BEC tran disp eq low fr zero cos}) and below.

\begin{figure}
\includegraphics[width=8cm,angle=0]{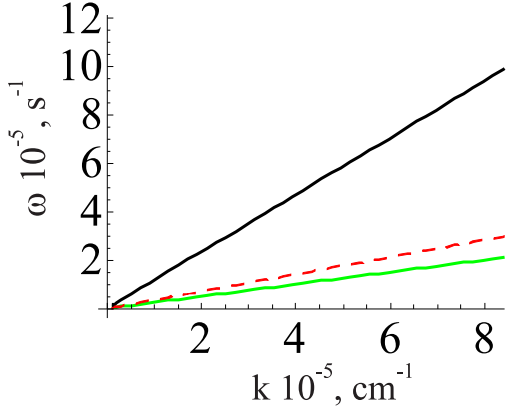}
\caption{\label{di BEC tran 02_disp_Bogol} (Color online) This figure shows dispersion of waves in electric dipolar BECs with evolution of dipole direction. Black (upper) line presents the dispersion dependence of the Bogoliubov mode (\ref{di BEC tran disp Bogol}). Green (lower) line shows dispersion of wave caused by evolution of the polarization $\textbf{P}$ (\ref{di BEC tran disp add for Bogol}). Dashed line shows the Bogoliubov mode (\ref{di BEC tran disp Bogol}) at zero equilibrium polarization $P_{0}=0$. This figure is obtained for $n_{0}=10^{12}$cm$^{-3}$, $d=1$D$=1$ $10^{-18}$CGS units, $m=10^{-23}$g, $a=10$nm, $\theta=\pi/4$.}
\end{figure}

Analysis of formula (\ref{di BEC tran disp dep low FR anizotr evident form}) was presented in Ref. \cite{Andreev arxiv 12 02} for $\textbf{k}\parallel \textbf{E}_{ext}$. This analysis includes contribution of the short range interaction up to the third order by the interaction radius terms proportional $\Upsilon_{2}$. So let us discuss some properties of formula (\ref{di BEC tran disp dep low FR anizotr evident form}) in the first order by the interaction radius of the short-range interaction (neglecting terms proportional $\Upsilon_{2}$). We also assume $\sigma=1$ and use $g=-\Upsilon$. Applying all this approximations we obtain two separate formulas from formula (\ref{di BEC tran disp dep low FR anizotr evident form}):
\begin{equation}\label{di BEC tran disp add for Bogol} \omega^{2}=\frac{1}{2m}gn_{0}k^{2}+\frac{\hbar^{2}k^{4}}{4m^{2}},\end{equation}
and
\begin{equation}\label{di BEC tran disp Bogol} \omega^{2}=\frac{1}{m}gn_{0}k^{2}+\frac{4\pi}{mn_{0}}P_{0}^{2}\cos^{2}\theta k^{2}+\frac{\hbar^{2}k^{4}}{4m^{2}}. \end{equation}
Assuming $P_{0}=dn_{0}$ we can rewrite dipolar part of spectrum as $\frac{4\pi d^{2}n_{0}k^{2}}{m}\cos^{2}\theta$.
Formula (\ref{di BEC tran disp Bogol}) is a generalization of the Bogoliubov spectrum for dipolar particles. Extra solution (\ref{di BEC tran disp add for Bogol}) appears as a consequence of polarization evolution. Solutions (\ref{di BEC tran disp add for Bogol}) and (\ref{di BEC tran disp Bogol}) are presented on Fig. (\ref{di BEC tran 02_disp_Bogol}).

Solution (\ref{di BEC tran disp Bogol}) is rather different from the well-known result containing signature of the roton instability (see for instance formula (5.1) in Ref. \cite{Lahaye RPP 09}, formula next to formula (6) in Ref. \cite{Baranov P R 08}, formula (11) in Ref. \cite{Baranov CR 12}, and formula (416) in Ref. \cite{Kawaguchi Ph Rep 12}). Experimental results in Ref. \cite{Bismut PRL 12} were obtained for the magnetic dipolar BECs. Consequently our analysis does not relate to this experiment. However we should admit that our previous results for the fully polarized BECs based on the full potential of magnetic dipole interaction gives an angle dependence of the spectrum corresponding to one obtained experimentally in Ref. \cite{Bismut PRL 12}. Hence our result (\ref{di BEC tran disp Bogol}) does not contradict to presence of the roton instability in spectrum of magnetic dipolar BECs. The Bogoliubov spectrum for the dipolar BECs presented in Refs. \cite{Lahaye RPP 09}, \cite{Baranov P R 08}, \cite{Baranov CR 12}, \cite{Kawaguchi Ph Rep 12} has the following form
\begin{equation}\label{di BEC tran disp from Lit}\omega^{2}=k^{2}\Biggl(\frac{n_{0}}{m}\biggl(g+\frac{C_{dd}}{3}(3\cos^{2}\theta-1)\biggr)+\frac{\hbar^{2}k^{2}}{4m^{2}}\Biggr),\end{equation}
$C_{dd}$ is the dipolar coupling
constant $C_{dd}=d^{2}/\varepsilon_{0}$ in the SI units, where $\varepsilon_{0}$ is the vacuum
permittivity, or $C_{dd}=4\pi d^{2}$ in the CGS units.
It has been applied to both the magnetic and the electric dipolar BECs. In our recent papers we shown that generalization of the Bogoliubov spectrum are different for the electric and the magnetic dipolar BECs even for the fully polarized BECs \cite{Andreev 1401.4842}, \cite{Andreev 2013 non-int GP}. These generalizations differ from formula (\ref{di BEC tran disp from Lit}). These differences appear due to the consideration of the full potentials of dipole-dipole interactions, which contain the Dirac delta function term (see formulas (\ref{di BEC tran Ham dd int}) and (\ref{di BEC tran togdestvo}) for the electric dipoles). Coefficients before the delta functions are different for the electric and magnetic dipole interactions. The delta function term in the electric dipole interaction gives an additional repulsion, when the delta function term in the magnetic dipole interaction gives an extra attraction (see also discussions in Ref. \cite{Andreev 1401.4842}). Nevertheless formula (\ref{di BEC tran disp Bogol}) corresponds to the result obtained for the fully polarized electric dipolar BECs \cite{Andreev 1401.4842}, \cite{Andreev 2013 non-int GP}, where the full potential of dipole-dipole interaction (\ref{di BEC tran Ham dd int}) was applied.

At $\cos\theta=0$ equation (\ref{di BEC tran disp eq low fr gen}) appears as
\begin{equation}\label{di BEC tran disp eq low fr zero cos} k^{2}c^{2}-\beta(\omega)\omega^{2}=0.\end{equation}
We did not have such limit case as equation (\ref{di BEC tran disp eq low fr zero cos}) when we considered longitudinal waves \cite{Andreev arxiv 12 02}, \cite{Andreev arxiv Pol}, \cite{Andreev RPJ 12}. Appearance of equation (\ref{di BEC tran disp eq low fr zero cos}) is related to consideration of the full set of Maxwell equations (\ref{di BEC tran M1})-(\ref{di BEC tran M3}) and presence of the perturbation of the electric field $\textbf{E}_{\perp}$ perpendicular to the direction of wave propagation. As a signature of presence of the transverse electric field $\textbf{E}_{\perp}$ we have the light speed in equation (\ref{di BEC tran disp eq low fr zero cos}) and its solutions below.

Explicit form of equation (\ref{di BEC tran disp eq low fr zero cos}) is
$$\Biggl(1+\frac{4\pi\sigma P_{0}^{2}}{mn_{0}c^{2}}\Biggr)\omega^{4}$$
$$-\biggl[\frac{\hbar^{2}k^{4}}{2m^{2}}+\frac{3}{2}\frac{gn_{0}k^{2}}{m}+\frac{\Upsilon_{2}n_{0}k^{4}}{m}+\frac{1}{2}\frac{4\pi P_{0}^{2}}{mn_{0}c^{2}}\frac{gn_{0}k^{2}}{m}$$
$$+\frac{4\pi\sigma P_{0}^{2}}{mn_{0}c^{2}}\Biggl(\frac{\hbar^{2}k^{4}}{4m^{2}}+\frac{gn_{0}k^{2}}{m}+\frac{\Upsilon_{2}n_{0}k^{4}}{m}\Biggr)\biggr]\omega^{2}$$
\begin{equation}\label{di BEC tran disp eq low fr zero cos explicit} +\Biggl(\frac{\hbar^{2}k^{4}}{4m^{2}}+\frac{gn_{0}k^{2}}{2m}\Biggr)\Biggl(\frac{\hbar^{2}k^{4}}{4m^{2}}+\frac{gn_{0}k^{2}}{m}+\frac{\Upsilon_{2}n_{0}k^{4}}{m}\Biggr)=0. \end{equation}
Neglecting contributions of the quantum Bohm potential (terms proportional $\hbar^{2}$) and the third order by the interaction radius (terms proportional $\Upsilon_{2}$) we find the following solution
$$\omega^{2}=\frac{gn_{0}k^{2}}{4m}\frac{1}{1+\frac{4\pi\sigma P_{0}^{2}}{mn_{0}c^{2}}}\times$$
\begin{equation}\label{di BEC tran disp eq low fr zero cos explicit solution} \times\Biggl[3+(2\sigma+1)\frac{4\pi P_{0}^{2}}{mn_{0}c^{2}}\pm\sqrt{\biggl(1+(2\sigma+1)\frac{4\pi P_{0}^{2}}{mn_{0}c^{2}}\biggr)^{2}+\frac{16\pi P_{0}^{2}}{mn_{0}c^{2}}}\Biggr].\end{equation}
Let us admit that solution (\ref{di BEC tran disp eq low fr zero cos explicit solution}) is obtained for $\cos\theta=0$. If $P_{0}\simeq0$ formula (\ref{di BEC tran disp eq low fr zero cos explicit solution}) simplifies to $\omega^{2}=\frac{gn_{0}k^{2}}{m}$ and $\omega^{2}=\frac{gn_{0}k^{2}}{2m}$, the first of them corresponds to the Bogoliubov spectrum. Getting the first order amendment on $\frac{4\pi P_{0}^{2}}{mn_{0}c^{2}}$, in the case of small polarization, we obtain
\begin{equation}\label{di BEC tran P1 B} \omega^{2}=\frac{gn_{0}k^{2}}{m}\biggl(1+\frac{4\pi P_{0}^{2}}{mn_{0}c^{2}}\biggr), \end{equation}
and
\begin{equation}\label{di BEC tran P1 N} \omega^{2}=\frac{gn_{0}k^{2}}{2m}\biggl(1-(1+\sigma)\frac{4\pi P_{0}^{2}}{mn_{0}c^{2}}\biggr). \end{equation}
In opposite limit, when $\frac{4\pi\sigma P_{0}^{2}}{mn_{0}c^{2}}\gg1$, we find $\omega^{2}=\frac{gn_{0}k^{2}}{m}\biggl(1+\frac{1}{2\sigma}\biggr)$ (modified Bogoliubov mode) and
\begin{equation}\label{di BEC tran P2}\omega^{2}=-\sigma\frac{gn_{0}k^{2}}{2m}\biggl(\frac{4\pi\sigma P_{0}^{2}}{mn_{0}c^{2}}\biggr)^{-1}.\end{equation}
The last solution has module of frequency much smaller than frequencies of the Bogoliubov mode. Solution (\ref{di BEC tran P2}) shows an aperiodic damping and no wave solution since $\omega^{2}<0$. On the other hand it might cause instability of dipolar BEC cloud to collapse.

\subsection{Electromagnetic waves}

In this subsection we do not use assumption of small frequencies. On the contrary we focus our attention on properties of high frequency excitations. If $P_{0}=0$ formula (\ref{di BEC tran disp eq}) gives one electromagnetic wave solution only: $\omega=kc$. In this case we do not have influence of the medium on properties of the electromagnetic waves. Let us admit that the model under consideration does not account resonance interaction of the electromagnetic waves with medium when frequencies of the waves cause electron transitions within atoms.

Under the condition $\theta=0 $ or $\theta=\pi$ equation (\ref{di BEC tran disp eq}) assumes the form
\begin{equation}\label{di BEC tran disp eq 0} (\omega^{2}-k^{2}c^{2})^{2}(1+\beta(\omega))=0. \end{equation}
From this equation we find usual the dispersion dependence of the light $\omega=kc$, and equation $1+\beta(\omega)=0$ describes the two matter waves, which appear instead of the Bogoliubov mode in the electrically polarized BECs, see formulas (\ref{di BEC tran disp dep low FR anizotr evident form})-(\ref{di BEC tran disp Bogol}) of this paper and Refs. \cite{Andreev arxiv 12 02}, \cite{Andreev arxiv Pol}, \cite{Andreev RPJ 12}.

In the opposite case, when $\theta=\pi/2$ equation (\ref{di BEC tran disp eq}) simplifies to
\begin{equation}\label{di BEC tran disp eq 0.5pi}(\omega^{2}-k^{2}c^{2})\biggl(\omega^{2}-k^{2}c^{2}+\beta(\omega)\omega^{2}\biggr)=0.\end{equation}
In this case we have two independent equation. One of them $\omega^{2}-k^{2}c^{2}=0$ describes changeless dispersion of the light. The second equation
\begin{equation}\label{di BEC tran disp eq 0.5pi lim of light}\omega^{2}(1+\beta(\omega))-k^{2}c^{2}=0\end{equation}
is the equation of the third degree of $\omega^{2}$. So, it contains three wave branches. We can admit that it contains both modified by the BEC dispersion of the light and the two matter wave. This equation has no low frequencies limitation as (\ref{di BEC tran disp eq low fr}), thus we have here high frequencies "tail" in the dispersion of the matter waves.

Let us consider influence of the BEC on the light as a small correction. Therefore we put $\omega=kc$ in $\beta(\omega)$ of equation (\ref{di BEC tran disp eq 0.5pi lim of light}), and accounting that $kc>>\hbar k^{2}/m$, $\sqrt{\Upsilon n_{0}/m}k$, $\sqrt{\Upsilon_{2}n_{0}/m}k^{2}$, in the result we find
\begin{equation}\label{di BEC tran light changed}\omega\simeq kc/\sqrt{1+\beta(kc)},\end{equation}
where $\beta(kc)=-4\pi n_{0}d^{2}(\frac{\sigma}{m}-\frac{gn_{0}}{2m^{2}c^{2}})/c^{2}$.
Expanding formula (\ref{di BEC tran light changed}) in a series by $\beta(kc)$ we get
\begin{equation}\omega=kc\biggl[1+\frac{2\pi n_{0}d^{2}}{mc^{2}}\biggl(\sigma-\frac{gn_{0}}{2mc^{2}}\biggr)\biggr],\end{equation}
with $P_{0}=n_{0}d$. We can see that the polarized BEC caused splitting of the light on two waves at light propagation perpendicular to the direction of the equilibrium polarization.
Magnitude of splitting $M_{s}$ is equal to
\begin{equation}\label{di BEC tran splitting} M_{s}\simeq kc\frac{2\pi\sigma n_{0}d^{2}}{m c^{2}}.\end{equation}
From this formula we find $M_{s}>0$ so frequency of the second wave is increased in compare with the $kc$: $\omega >kc$. However this shift is very small $M_{s}\simeq 10^{-18}kc$ for $n_{0}=10^{15}$cm$^{-3}$, $d=3$ $10^{-18}$ CGS units, $m=10^{-23}$ g.

In the large frequency limit we can put $\beta=\beta(\omega=kc)$ in formula (\ref{di BEC tran disp eq}). When formula (\ref{di BEC tran disp eq}) becomes an equation of the second degree for $\omega^{2}$. Solving this approximate equation for all $\theta$ we get two solutions
\begin{equation}\label{di BEC tran}\omega^{2}=k^{2}c^{2},\end{equation}
and
\begin{equation}\label{di BEC tran}\omega^{2}=\frac{1+\beta\cos^{2}\theta}{1+\beta}k^{2}c^{2},\end{equation}
accumulating all results described in this subsection.

\section{\label{sec:level1} Conclusion}

We have considered transverse waves in the electrically polarized BEC. To do it we have used whole set of the Maxwell equation for description of electromagnetic field caused by the dynamic of electric dipoles of the medium. In comparison to the case of longitudinal waves considered in our previous papers we have found an advantage, which is a possibility of consideration of the light propagating through the electrically polarized BECs. Thus, we have considered dispersion of the light propagating through the medium and we have found the splitting of the light on two waves in the case when the light propagate at the angle to the direction of the equilibrium polarization. We have calculated the magnitude of frequencies splitting for the case of the light propagation perpendicular to the direction of equilibrium polarization. There is no splitting in the case of the light propagation along the direction of the  equilibrium polarization, there is also no contribution of the medium in the light dispersion dependence in this case.

As in the case of longitudinal waves we have got two matter waves, which exist instead of the Bogoliubov mode existing in the unpolarized BECs and fully polarized BECs. We have found analytical solution for the plane matter waves propagating in the three dimensional BECs and shown the influence of the anisotropy on the value of the frequencies.

In the result, we have developed generalization of the quantum hydrodynamic equations for the electrically polarized BECs including transverse wave propagation. Using it we have shown that in the electrically polarized BEC situated in the external electric field there are four wave, two of them are high frequencies waves, and them associated with the light. Two more wave are the anisotropic matter waves.

\section{\label{sec:level1} Appendix A: potential of electric dipole interaction}

The potential of electric field created by a single electric dipole $\textbf{d}_{i}$ appears as
\begin{equation}\label{di BEC tran pot of el field} \varphi(\textbf{r})=-(\textbf{d}_{i}\nabla)\frac{1}{r}\end{equation}
(see for instance \cite{Landau 2}).
We can take the derivative in formula (\ref{di BEC tran pot of el field}) and obtain more explicit form of $\varphi(\textbf{r})$. However it is more illuminating to keep $\varphi(\textbf{r})$ in form (\ref{di BEC tran pot of el field}). The potential energy of an electric dipole $\textbf{d}_{j}$ in the electric field of another dipole $\textbf{d}_{i}$ is $U_{dd}=-\textbf{d}_{j}\textbf{E}_{ji}$, where $\textbf{E}_{ji}$ is the electric field created by the electric dipole $\textbf{d}_{i}$ in the position of the dipole $\textbf{d}_{j}$. Using (\ref{di BEC tran pot of el field}) we find \begin{equation}\label{di BEC tran el field of a dipole}\textbf{E}_{ji}=-\nabla\varphi=(\textbf{d}_{i}\nabla)\nabla\frac{1}{r}.\end{equation}
In tensor notations it can be presented as $E^{\alpha}_{ji}=d^{\beta}_{i}\nabla^{\beta}\nabla^{\alpha}(1/r)$. Let us have a look on the differential operator in formula (\ref{di BEC tran el field of a dipole}). It is a tensor operator emerges as $\nabla^{\beta}\nabla^{\alpha}(1/r)$. The explicit form of this operator is presented by formula (\ref{di BEC tran togdestvo}). We may now apply formula (\ref{di BEC tran togdestvo}) to get explicit form of electric field created by dipoles. As the result we find difference with usually used formula (\ref{di BEC tran pot of dd int usual}).

Real atoms and molecules have a finite size, which can be presented by hard-core short range potential. It leads to cutting off of potential of the dipole-dipole interaction at small interparticle distances
\begin{equation}\label{di BEC tran}U_{int}=\biggl[\begin{array}{ccc} U_{ij,SR}+U_{ij,LR}&
for &
r>r_{0}\\
+\infty &
for &
r\leq r_{0}, \\
\end{array}\end{equation}
where $U_{ij,LR}$ is the potential of the long range dipole-dipole interaction, and $U_{ij,SR}$ is the potential of the short range interaction presenting a field surrounding a neutral particle. Potential is short range means that $U_{ij,SR}\rightarrow 0$ at $r\sim(2\div3)r_{0}$. See Ref. \cite{Andreev 1401.4842} for the Bogoliubov spectrum of the finite size dipolar BECs.

For non-dipolar BECs we have
\begin{equation}\label{di BEC tran}U_{int}=\biggl[\begin{array}{ccc} U_{ij,SR}&
for &
r>r_{0}\\
+\infty &
for &
r\leq r_{0}. \\
\end{array}\end{equation}
So we have the following constant of the short range interaction for the cut off potential $g=4\pi\int_{r_{0}}^{+\infty}U_{SR}(r)r^{2}dr$.

\section{\label{sec:level1} Appendix B: definition of quantum hydrodynamic variables}

The concentration of particles is defined as the quantum average of the concentration operator in the coordinate representation $\hat{n}=\sum_{i}\delta(\textbf{r}-\textbf{r}_{i})$ with the sum over all particles in the system:
\begin{equation}\label{di BEC tran def density}n(\textbf{r},t)=\int dR\sum_{i}\Psi^{*}(R,t)\delta(\textbf{r}-\textbf{r}_{i})\Psi(R,t)\end{equation}
where $R=\{\textbf{r}_{1},\textbf{r}_{2},...,\textbf{r}_{N}\}$ is a vector in $3N$ dimensional configurational space containing set of coordinates of all $N$ particles in the system under consideration, $dR=\prod_{p=1}^{N}d\textbf{r}_{p}$ is the volume element in $3N$ dimensional space, $\Psi(R,t)$ is the wave function describing exact evolution of $N$ particle quantum system. $\Psi(R,t)$ obeys the many-particle Schrodinger equation. The operator of particle concentration contains the Dirac delta functions, which provide projection of $3N$ dimensional quantum dynamics in three dimensional physical space, where real particles move.

The particle flux or the momentum density appears as
$$j^{\alpha}(\textbf{r},t)=\int dR\sum_{i}\delta(\textbf{r}-\textbf{r}_{i})\frac{1}{2m_{i}}\times$$
\begin{equation}\label{di BEC tran def of current of density}\times\biggl(\Psi^{*}(R,t)p_{i}^{\alpha}\Psi(R,t)+c.c.\biggr),\end{equation}
where $\textbf{p}_{i}=-\imath\hbar\nabla_{i}$ is the momentum operator of the $i$th particle, and $c.c.$ stands for the complex conjugation.
The particle flux allows to derive the Euler equation applying $\textbf{j}=n\textbf{v}$.

Polarization appears in the Euler equation as follows
\begin{equation}\label{di BEC tran def polarization}P^{\alpha}(\textbf{r},t)=\int dR\sum_{i}\delta(\textbf{r}-\textbf{r}_{i})\Psi^{*}(R,t)\hat{d}_{i}^{\alpha}\Psi(R,t),\end{equation}
where $\hat{d}_{i}^{\alpha}$ is the operator of electric dipole moment of a neutral particle.

The polarization current or polarization flux is
$$R^{\alpha\beta}(\textbf{r},t)=\int dR\sum_{i}\delta(\textbf{r}-\textbf{r}_{i})\frac{\hat{d}_{i}^{\alpha}}{2m_{i}}\times$$
\begin{equation}\label{di BEC tran def of current of polarization} \times\biggl(\Psi^{*}(R,t)p_{i}^{\beta}\Psi(R,t)+c.c.\biggr).\end{equation}

The internal electric field caused by dipoles has the following explicit form
\begin{equation}\label{di BEC tran el field of dip explict} E^{\alpha}(\textbf{r},t)=\int d\textbf{r}' G^{\alpha\beta}(\textbf{r},\textbf{r}')P^{\beta}(\textbf{r}',t).\end{equation}

To derive QHD equations we need the Schrodinger equation $\imath\hbar\partial_{t}\Psi=\hat{H}\Psi$, which governs evolution of the many-particle wave function. We also need explicit form of the Hamiltonian of the system under consideration. Which within the
quasi-static approximation is
$$\hat{H}=\sum_{i}\Biggl(\frac{1}{2m_{i}}\hat{\textbf{p}}_{i}^{2}-\widehat{\textbf{d}}_{i}\textbf{E}_{i,ext}+V_{trap}(\textbf{r}_{i},t)\Biggr)$$
\begin{equation}\label{di BEC tran Hamiltonian} +\frac{1}{2}\sum_{i,j\neq i}\Biggl(U_{ij}-\hat{d}_{i}^{\alpha}\hat{d}_{j}^{\beta}G_{ij}^{\alpha\beta}\Biggr).\end{equation}
The first term in the Hamiltonian is the operator of the kinetic energy. The second
term represents the interaction between the dipole moment
$\hat{d}_{i}^{\alpha}$ and the external electric field. The subsequent
terms represent the short-range $U_{ij}$ and the long range dipole-dipole $d_{i}^{\alpha}d_{j}^{\beta}G_{ij}^{\alpha\beta}$
interactions between neutral particles baring electric dipole moment. The Green function
for the dipole-dipole interaction reads as
$G_{ij}^{\alpha\beta}=\nabla^{\alpha}_{i}\nabla^{\beta}_{i}(1/r_{ij})$.

At derivation of the QHD equations we use quasi static approximation for the dipole-dipole interaction. It means that at each moment of time we consider dipole-dipole interaction between motionless dipoles. This interaction changes state of the translational motion of molecules and direction of their dipoles. At next moment of time particles are in new positions and having new direction of dipoles. In this new state particles interact as motionless, etc. Quasi static interaction is an interaction  via the electric field created by motionless dipoles. Slow motion of dipoles creates small magnetic field and electromagnetic radiation. Which can be accounted by consideration of the full set of Maxwell equations, as we do it in the end of section II.

More details on derivation of quantum collective observable evolution by quantum hydrodynamic method can be found in the following Refs.  \cite{Andreev PRB 11}, \cite{MaksimovTMP 1999} (focused on dipole evolution), \cite{Andreev PRA08} (focused on BECs evolution).


\end{document}